\documentclass[12pt]{article}
\usepackage{epsfig}
\def\be{\begin{eqnarray}}
\def\ee{\end{eqnarray}}

\begin{document}
\begin{titlepage}
\title{Asymmetric Nuclear Matter from  Extended 
Brueckner-Hartree-Fock Approach}
\author{\normalsize W. Zuo$^{1}$ 
\footnote{Permanent address: Institute of Modern Physics, Lanzhou, China}, 
I. Bombaci$^{2,3}$, U. Lombardo$^{1,4}$ \\ \normalsize
$^{1}$INFN-LNS, 44 Via S.~Sofia, I-95123 Catania, Italy \\ \normalsize
$^{2}$ Dipartimento di Fisica, Universit\`a di Pisa, via Buonarroti, 2,
56127 Pisa, Italy \\ \normalsize
$^{3}$ I.N.F.N., Sezione di Pisa,  via Buonarroti, 2, 56127 Pisa, Italy \\ 
\normalsize
$^{4}$ Dipartimento di Fisica, 57 Corso Italia, I-95129 Catania, Italy }
\maketitle
\vskip .2 truecm
\centerline{\bf Abstract}
The properties of isospin-asymmetric nuclear matter have been 
investigated in the framework of the extended Brueckner-Hartree-Fock 
approximation at zero temperature. Self-consistent calculations using 
the Argonne $V_{14}$ interaction are reported for several asymmetry 
parameters $\beta = \frac{N - Z}{A}$ ranging from symmetric 
nuclear matter to pure neutron matter. The binding energy per nucleon 
fulfills the $\beta^2$ law in the whole asymmetry range. The symmetry 
energy is calculated for different densities and discussed in 
comparison with other predictions. At the saturation point it is in 
fairly good agreement with the empirical value. The present 
approximation, based on the Landau definition of quasiparticle energy, 
is investigated in terms of the Hugenholtz-Van Hove theorem, which 
is proved to be fulfilled with a good accuracy at various asymmetries. 
The isospin dependence of the single-particle properties is discussed, 
including mean field, effective mass, and mean free path of neutrons 
and protons. The isospin effects in nuclear physics and nuclear 
astrophysics are briefly discussed. \\[.2cm]
\end{titlepage}

\section{Introduction}

Within the general interest for the equation of state ( EOS ) of 
nuclear matter in nuclear physics as well as in nuclear astrophysics, 
increasing attention is currently paid to the isospin degree of freedom. 

The EOS of isospin asymmetric nuclear matter plays a central role for our
understanding of astrophysical phenomena like supernova explosions, neutron
stars structure, X-ray bursts, neutron stars merging and possibly
$\gamma$-ray bursts.
The study of asymmetric nuclear matter represents also the first step for a
microscopic theory of the structure of nuclei far from the valley of
beta stability. This ``terra incognita'' is going to be explored in the 
near future thanks to a new generation of experimental facilities with 
high intensity radioactive ion beams. 
Moreover, dynamical simulations of collisions between neutron-rich nuclei 
show that the main reaction mechanisms including fragmentation are quite 
sensitive to the density dependence of the nuclear 
symmetry energy \cite{BAO,DITO}. 
Such calculations mainly make use of phenomenological Skyrme-like forces 
where the symmetry energy at high density can also be in strong 
disagreement with the one extracted from the microscopic predictions. 
\par
On a microscopic basis the EOS of asymmetric nuclear matter 
has been studied within the variational approach \cite{WFF,PANDA,MODA} 
as well as relativistic \cite{MALF,ENG,HUB,KUO,FRO,JONG} and 
non-relativistic \cite{LOM} Brueckner-Bethe-Goldstone (BBG) theory. 
Within the Brueckner-Hartree-Fock (BHF) approximation to the BBG 
theory a systematic study of isospin effects on the EOS of asymmetric 
nuclear matter has been carried out in Ref.~\cite{LOM}, where 
a separable version \cite{HEIDE} of the Paris potential \cite{PARI} was 
adopted to describe the two-body nuclear force.

Beside the bulk properties (EOS), the authors of Ref.~\cite{LOM} 
focused also on the single-particle (s.p.) properties 
of neutrons and protons in isospin-asymmetric
nuclear medium. The neutron and proton s.p. potentials were 
calculated \cite{LOM} to the lowest order in the Brueckner reaction 
matrix (BHF approximation), using 
the so-called {\it continuous choice}~\cite{MAHA0}. 

Motivated by the renewed interest in this subject, in the present 
paper, we report an extension of the calculations of 
Ref.~\cite{LOM} along the following lines. First, in the calculations 
we make use of a different realistic nucleon-nucleon ( $NN$ ) 
potential, {\it i.e.,} the full Argonne $V_{14}$
potential \cite{WIRI}, which enables us to take into account a larger
number of partial waves with respect to the calculation \cite{LOM} 
with the separable Paris potential. These additional partial 
waves ( $3\le L \le 6$ ) give a non-negligible contribution both to 
the EOS and the nucleon mean field, expecially in the high-density 
region, which is relevant for applications in astrophysics as well 
as heavy ion physics.

Second, the Bethe-Goldstone equation is now solved for the complex 
$G$-matrix. This enables us to calculate the complex nuclear mean 
field and some closely related quantities such as the optical 
potential and the mean free path.
   
Third, according to the Landau definition of quasi-particle 
energy (for an extended discussion see Ref.~\cite{BROW0}) in the 
calculations of the mass operator (nucleon self-energy) and 
single-particle properties, we go beyond the BHF approximation by 
including some higher-order correlation contributions. 
In particular, in the present work, we include the so-called 
{\it rearrangement} term $M_2$ which is a second order diagram in the 
$G$-matrix and accounts for particle-hole excitations in nuclear 
matter ground state. Next we consider also 
the {\it renormalization} contributions of the third and forth 
order in the $G$-matrix, which account for the partial depletion 
of the neutron and proton Fermi seas due to the nuclear 
correlations~\cite{MAHA}. It has been shown, in the case of pure 
neutron matter \cite{ZUO1} and also symmetric nuclear 
matter~\cite{ZUO2} that the new terms give a large contribution to 
s.p. properties like the mean field and the nucleon effective 
mass. We will refer to the present approach to compute nuclear 
s.p. properties as extended Brueckner--Hartree--Fock ( EBHF ) 
approximation \cite{ZUO1,ZUO2}.

As is well known, the BHF approximation largely violates the 
Hugenholtz-Van Hove ( HVH ) theorem \cite{HVH}, which basically 
measures the consistency of a given order of approximation in a 
perturbative approach. In symmetric nuclear matter, the inclusion of 
the rearrangement contribution greatly improves the fulfillment of 
the HVH theorem \cite{BAL1}. In the present paper, we study this problem 
in the case of asymmetric nuclear matter within the EBHF approximation. 

\section{EBHF and Nucleon Self-energy for Asymmetric Nuclear Matter}

In this section the formalism of the Brueckner-Bethe-Goldstone ( BBG ) 
theory is described for the case of asymmetry nuclear 
matter \cite{LOM,BKL}. The proton and the neutron Fermi momenta are 
related to their corresponding densities 
$\rho_{\rm p}$ and $\rho_{\rm n}$ through the relations 
\be
\nonumber
k_F^{\rm p} = [\frac{3\pi^2}{2}(1-\beta)\rho]^{1/3} ,
\cr\noalign{\vskip 3mm}
k_F^{\rm n} = [\frac{3\pi^2}{2}(1+\beta)\rho]^{1/3} ,
\ee
where $\rho=\rho_{\rm p}+\rho_{\rm n}$ is the total density, and 
$\beta=(\rho_{\rm n}-\rho_{\rm p}) / \rho$ the asymmetry 
parameter determining the neutron excess (from now on we assume 
$ \rho_{\rm n} \ge \rho_{\rm p}$).  

The starting point in BBG theory, is the Brueckner reaction matrix $G$,  
which in the case of asymmetric nuclear matter depends also on the 
isospin components of the two colliding nucleons.  
The $G$-matrix satisfies the Bethe-Goldstone equation, 
\be
G(\rho,\beta;\omega) =  v_{NN} + v_{NN}  \sum_{k_1 k_2}
 \frac { |k_1 k_2 \rangle Q(k_1,k_2) \langle k_1 k_2| }
{\omega - \epsilon(k_1)-\epsilon(k_2)+i\eta}
G(\rho,\beta;\omega) \ ,
\label{eq:BG}
\ee
where $v_{NN}$ is the two-body nuclear interaction and $\omega$ the 
starting energy. Here $k\equiv (\vec k,\sigma,\tau)$ denotes s.p. 
momentum, $z$-components of spin and isospin, respectively. 

The $G$-matrix can be considered as an in-medium effective interaction 
between two nucleons. The surrounding nucleons renormalize the bare 
$NN$ interaction via the Pauli blocking and the nuclear mean field. 
The Pauli operator, defined as 
\be
Q(k_1,k_2) = [1-n(k_1)] [1-n(k_2)] \ , 
\label{eq:pauli}
\ee
prevents two nucleons in intermediate states from scattering into 
states inside their respective Fermi seas. By $n(k)$ we denote the 
Fermi distribution function, which at zero temperature is given by 
the step function $\theta(k-k_F^\tau)$ ( uncorrelated ground state ). 
The s. p. energy 
%%%%%%%%%%%%%%%%%%%%%%%%%%%%%%%%%%%%%%%%%%%%%%%%%%%%%%%%%%
{\footnote{In the present work, we assume the neutron and 
proton rest masses equal to their average value $m$}}
%%%%%%%%%%%%%%%%%%%%%%%%%%%%%%%%%%%%%%%%%%%%%%%%%%%%%%%%%%
\be
\epsilon(k) = \frac{\hbar^2k^2}{2m}  +  U(k) \ , 
\label{eq:e1}
\ee
appearing in the energy denominator of Eq.~(\ref{eq:BG}), 
involves the {\it auxiliary} potential $U(k)$, which controls 
the convergence rate of the hole-line expansion. Within the BHF 
approximation the neutron and proton s.p. auxiliary potentials 
are calculated from the real part of the on-shell 
antisymmetrized $G$-matrix, via the relation 
\be
 U(k) = 
\sum_{k'} n(k') 
{\rm Re} 
    \langle k k'|G(\epsilon(k) + \epsilon(k'))
                              |k k'\rangle_A \ . 
\label{eq:U1}
\ee
Here we adopt the continuous choice \cite{MAHA0} for the auxiliary s.p. 
potential. In this context, the auxiliary potential has the 
physical meaning of the mean field that each nucleon feels during 
its propagation between two successive scatterings.
 
In the BHF approximation, Eqs.~(\ref{eq:BG}),(\ref{eq:e1}), and 
(\ref{eq:U1}) are solved self-consistently for given total density 
$\rho$ and asymmetry $\beta$. 
Then the energy per particle is evaluated at the 
lowest order ( two hole-line diagrams ) of the BBG hole-line 
expansion ( see Ref.~\cite{BKL} for the case of asymmetric 
matter ). 
   
%%%%%%%%%%%%%%%%%%%%%%%%%%
\subsection{Mass operator and quasi-particle energy}
%%%%%%%%%%%%%%%%%%%%%%%%%%
One of the main purposes of the present paper is to calculate s.p. 
properties of neutrons and protons in asymmetric matter going beyond the 
BHF approximation.  
To this end we introduce the mass operator \cite{MAHA0,MS} 
\be
       M^\tau(k,\omega) = V^\tau(k,\omega) + i W^\tau(k,\omega) \ ,
\ee
which is a complex quantity and can be identified with the potential 
energy felt by a neutron~($\tau=n$) or a proton ($\tau=p$) 
with momentum $\vec k$ 
and energy $\omega$ in asymmetric nuclear matter ( hereafter, we will 
write out explicitly the isospin index $\tau$ ). 
In the same spirit of the BBG theory, the mass operator 
$M^\tau(k,\omega)$ can be expanded in a perturbation series 
according to the number of hole lines \cite{HM} and various terms of 
this expansion can be represented by means of Goldstone diagrams 
a few of which are shown in Fig.~1. 

\begin{figure}
\centerline{\epsfig{figure = 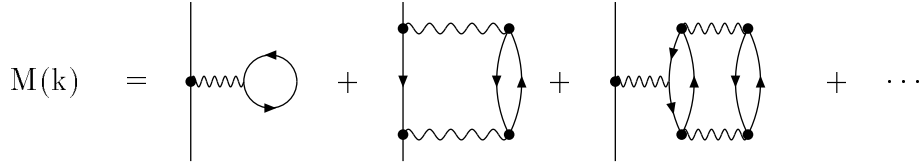}}
\caption{ \small{Hole-line expansion of s.p. potential.} }
\end{figure}
\normalsize

In analogy with the case of symmetric nuclear matter, the neutron and 
proton quasi-particle energies $E^{\tau}(k)$ are the solutions 
of the energy-momentum relation 
\be
E^\tau(k) = \frac{\hbar^2k^2}{2m} + V^\tau(k,E^\tau(k)) \ ,
\label{eq:qpe}
\ee
{\it i.e.,} $E^\tau(k)$ is obtained from the on-shell values of the real 
part of the mass operator.  

\begin{figure}[ht]
\centerline{\epsfig{figure = 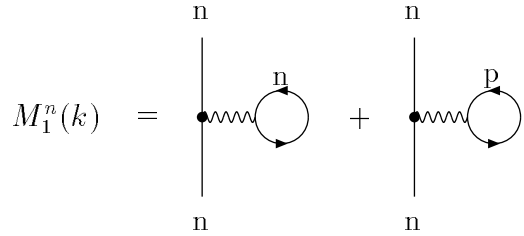}}
\caption{\small{The first order hole-line expansion of neutron 
s.p. potential. }}
\end{figure}
\normalsize

To the lowest order in the hole-line expansion the mass operator is 
given by ( diagrams of Fig.~2 ) 
\be
 M_1^\tau(k,\omega) = 
\sum_{\tau'} \sum_{\vec k' \sigma'} n^{\tau'}(k') 
  \langle k k'|G^{\tau\tau'}(\omega + \epsilon^{\tau'}(k'))|k k'\rangle_A 
\equiv      \sum_{\tau'} M_1^{\tau\tau'}(k,\omega) \ .  
\label{eq:M1}
\ee
In this approximation the quasi-particle energy $E_1^\tau(k)$ coincides  
with the BHF s.p. energy given by Eqs.~(\ref{eq:e1}) and 
(\ref{eq:U1}), i.e., $E_1^\tau(k) = \epsilon^{\tau}(k)$. 

%%%%%%%%%%%%%%%%%%%%%%%%%%
\subsection{The rearrangement contribution to the s.p. energy}
%%%%%%%%%%%%%%%%%%%%%%%%%%
The next contribution to the perturbative expansion of the mass operator 
is given by the so-called {\it rearrangement} term 
$M_2^\tau(k,\omega)$ \cite{MAHA0}. The associated Goldstone diagrams 
are shown in Fig.~3. $M_2^\tau$ is a second-order diagram in the 
$G$-matrix and accounts for particle-hole excitations in nuclear 
matter. Its expression, extended to asymmetric nuclear matter, reads 
\be
M_2^\tau(k,\omega) &=&   \frac{1}{2} \sum_{\tau'} \sum_{\vec k'\sigma'} 
(1-n^{\tau'}(k')) \sum_{k_1 k_2} n^{\tau}(k_1) n^{\tau'}(k_2)
\frac {|\langle k k'|
G^{\tau\tau'}(\epsilon^{\tau}(k_1)+\epsilon^{\tau'}(k_2))| 
                                                   k_1 k_2\rangle_A|^2}
{\omega + \epsilon^{\tau'}(k')   - \epsilon^{\tau}(k_1) 
        - \epsilon^{\tau'}(k_2) - i\eta}
\cr\noalign{\vskip 3mm} 
  &\equiv&  \sum_{\tau'} M_2^{\tau\tau'}(k,\omega) \ , 
\label{eq:M2}
\ee
where $\epsilon^{\tau}(k)$ is the s.p. spectrum in BHF approximation, 
given by Eqs.~(\ref{eq:e1}) and (\ref{eq:U1}).  
In this approximation for the mass operator 
[ {\it i.e.}, $M^\tau(k,\omega) \simeq M_1^\tau(k,\omega) 
+ M_2^\tau(k,\omega) $ ], the quasi-particle energy (\ref{eq:qpe}) 
is given by the approximate relation 
\be
E_{2}^\tau(k) &=& E_{1}^\tau(k) +  Z_{2}^\tau(k) V_2^\tau(k,E_{1}^\tau(k)) 
\cr\noalign{\vskip 3mm}
&=&  \frac{\hbar^2k^2}{2m} + V_1^\tau(k,E_1^{\tau}(k)) 
                            +  Z_{2}^\tau(k) V_2^\tau(k,E_{1}^\tau(k)) \ , 
\label{eq:qpe2}
\ee 
where 
\be
 Z_{2}^\tau(k) = \Bigg\{ 1 - \frac{\partial}{\partial \omega}
 \bigg[ V_1^\tau(k,\omega)+V_2^\tau(k,\omega) \bigg] 
\Bigg\}^{-1}_{\omega=E_1^\tau(k)} 
\ee
is an approximation of the {\it quasi-particle strength} for asymmetric 
nuclear matter
\be
 Z^\tau(k) = \Bigg\{ 1 - \frac{\partial}{\partial \omega}
 \bigg[ V^\tau(k,\omega)\bigg] \Bigg\}^{-1}_{\omega=E^\tau(k)} \ . 
\ee

\begin{figure}
\centerline{\epsfig{figure = 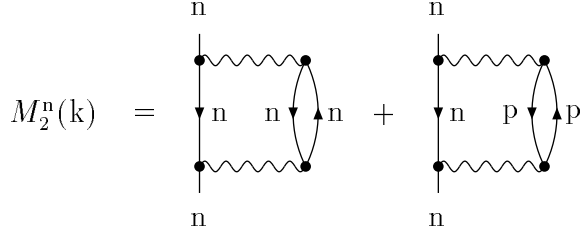}}
\caption{\small{The second order hole-line expansion 
of neutron s.p. potential. }}
\end{figure}
\normalsize

%%%%%%%%%%%%%%%%%%%%%%%%%%
\subsection{The renormalization contributions to the s.p. energy}
%%%%%%%%%%%%%%%%%%%%%%%%%%
Due to many-body correlations the two Fermi seas are partially depleted, 
and the correlated momentum distributions $\tilde{n}^\tau(k)$ 
differ from the uncorrelated ones $n^\tau(k)=\theta(k-k_F^\tau)$. 
To account for this physical effect, 
one considers the contribution $M^\tau_3(k,\omega)$ ( last diagram 
of Fig.~1 ) given by \cite{MAHA0,MAHA}  
\be
M_3^{\tau}(k,\omega)= 
- \sum_{\tau'}\sum_{\vec{h'} \sigma'}\kappa_2^{\tau'}(h')
\langle kh'|G^{\tau\tau'}(\omega + \epsilon^{\tau'}(h'))|kh'\rangle_A \ , 
\label{eq:M3}
\ee
where $h'$ refers to ``hole'' state with momentum smaller 
than $k_{\rm F}^\tau$, and 
\be
\nonumber
\kappa_2^{\tau'}(h')=-\left[
\frac{\partial}{\partial\omega}
M_1^{\tau'}(h',\omega)\right]_{\omega=\epsilon^{\tau'}(h')}
%\label{eq:kappa2}
\ee
is at the lowest order the depletion of neutron (proton) Fermi 
sea \cite{MAHA0,MAHA}, {\it i.e.,}  $\kappa_2^{\tau'}(h')$ is 
the probability that a neutron (proton) hole-state 
$(|\vec h'| \le k_F^{\tau'})$ is empty.  
Let us consider now the sum 
\be
 \widetilde{M}_1^\tau(k,\omega) &\equiv& 
 M_1^\tau(k,\omega) +  M_3^\tau(k,\omega) 
\cr\noalign{\vskip 3mm}
&=& 
\sum_{\tau'} \sum_{\vec h' \sigma'} \Big[ 1 - \kappa_2^{\tau'}(h')  \Big]
\langle k h'|G^{\tau\tau'}(\omega + \epsilon^{\tau'}(h'))|k h'\rangle_A 
\cr\noalign{\vskip 3mm}
&=& 
\sum_{\tau'} \sum_{\vec h' \sigma'}  \
\tilde{n}_2^{\tau'}(h') 
\langle k h'|G^{\tau\tau'}(\omega 
+ \epsilon^{\tau'}(h'))|k h'\rangle_A \ , 
\label{eq:M1ren}
\ee
$\tilde{n}^{\tau'}_2(h') = \Big[ 1 - \kappa_2^{\tau'}(h') \Big]$ 
being the second-order approximation for the correlated momentum 
distribution. $ \widetilde{M}_1^\tau(k,\omega)$ is 
the so-called {\it renormalized} BHF approximation for 
the off-shell mass operator [ compare to Eq.~(\ref{eq:M1}) ]. 

An accurate approximation consists in using the average value of the 
depletion, which is 
\be
\kappa^{\tau'} = 
        \kappa_2^{\tau'} \big(h'=0.75k^{\tau'}_{\rm F}\big) \ . 
\ee
Then Eqs.~(\ref{eq:M3}) and (\ref{eq:M1ren}) yield 
\be
M_3^\tau(k,\omega) \approx 
        - \sum_{\tau'} \kappa^{\tau'} M_1^{\tau\tau'}(k,\omega) \ , 
\label{eq:M3'}
\ee
\be
 \widetilde{M}_1^\tau(k,\omega) \approx 
\sum_{\tau'} \Big[ 1 - \kappa^{\tau'} \Big] M_1^{\tau\tau'}(k,\omega) \ . 
\label{eq:M1ren'}
\ee

From the similar considerations, a renormalization correction 
should also be brought from the four hole-line terms to the 
second-order contribution to $M_2^{\tau}$ in order to take 
into account the fact that the hole-state $k_1$ in 
Eq.~(\ref{eq:M2}) is partially 
empty ( see also Ref.~\cite{MAHA} for symmetric nuclear matter ). 
Along the same line of the previous correction one gets 
the {\it renormalized} $M_2$, which is approximately given by 
\be
\widetilde{M}_2^{\tau}(k,\omega)=
\sum_{\tau'} \Big[ 1 - \kappa^{\tau'} \Big] M_2^{\tau\tau'}(k,\omega) \ . 
\label{e:Uren2}
\ee
The renormalized contributions can also be traced to the functional 
dependence of the $G$-matrix on the quasi-particle occupation numbers 
within the Landau theory of Fermi liquids. It can be shown, in fact, 
that taking the functional derivative of the binding 
energy (at two hole-line level) includes also the terms 
of the third and fourth order in the self-energy, the effect of which 
has just been discussed. Taking into account all the corrections 
discussed above, from Eq.~(\ref{eq:qpe}) one can get the following 
expression for the quasi-particle energy \cite{MAHA}, 
\be
E^{\tau}(k) &\simeq& \frac{\hbar^2 k^2}
{2m} +
V_1^\tau({k},E_{1}^\tau(k))
\cr\noalign{\vskip 3mm}
&+& Z^\tau (k)\sum_{\tau'}  
\left[ -\kappa^{\tau'} V_1^{\tau\tau'}(k,E_{1}^{\tau}(k))
+ (1-\kappa^{\tau'})V_2^{\tau\tau'}(k,E_{1}^{\tau}(k))\right] \ , 
\label{e:Espec}
\ee
where 
\be
 Z_{3}^\tau(k) = \Bigg\{ 1 - \sum_{\tau'} ( 1- \kappa^{\tau'}) 
 \frac{\partial}{\partial \omega}
 \bigg[ V_1^{\tau\tau'}(k,\omega)+V_2^{\tau\tau'}(k,\omega) 
 \bigg] \Bigg\}^{-1}_{\omega=E_1^\tau(k)} \ . 
\ee
In the following we refer to this approximation for the 
quasi-particle energy as the extended 
Brueckner-Hartree-Fock (EBHF) approximation \cite{MAHA,ZUO1,ZUO2}.   

\noindent
\subsection{Partial wave expansion and angular averaging}
After the usual angular averaging on the Pauli operator and the 
energy denominator \cite{GRAN,BAL1}, the Bethe-Goldstone equation 
can be expanded in partial waves, 
\be
&  & G^{\tau\tau'}_{\alpha LL'}(q,q',P,\omega)
= v_{\alpha LL'}(q,q') 
\cr\noalign{\vskip 3mm}
&+& 
\frac{2}{\pi}
\sum_{L''}
\int q''^2 {\rm d}q'' v_{\alpha LL''}(q,q'')
\frac{\langle Q^{\tau\tau'}(q'',P) \rangle}
{\omega-e_{12}^{\tau\tau'}(q'',P)+i\eta}
G^{\tau\tau'}_{\alpha L''L'}(q'',q',P,\omega) \ , 
\ee
where $\vec{q}=(\vec{k}_1-\vec{k}_2)/2$ and 
$\vec{P}=\vec{k}_1+\vec{k}_2$ are the relative momentum and 
total momentum, respectively. $e_{12}^{\tau\tau'}(q'',P)= 
\langle\epsilon^{\tau}(k_1)+\epsilon^{\tau'}(k_2)\rangle$
is the angle average of the energy denominator. 
The angular-averaged Pauli operator is 
\par
  (i) for $\tau = \tau'$ (neutron-neutron or proton-proton), 
\be
\langle Q^{{\tau}{\tau}}(q,P)\rangle &=&
\left\{ 
\begin{array}{llc}
\displaystyle 
{\rm min}(1,\xi_{\tau})
& \quad &
{\rm if} \hspace*{.2cm} \xi_{\tau}\ge 0  \\
0 & & 
{\rm otherwise} \ , 
\end{array}
\right.  
\ee
\par
  (ii) for $\tau \ne \tau'$ (neutron-proton or proton-neutron), 
\be
 \langle Q^{{\tau}{\tau'}}(q,P)\rangle &=&
\left\{
\begin{array}{llc}
\displaystyle
\frac{1}{2}[{\rm min}(1,\xi_{\rm p})+{\rm min}(1,\xi_{\rm n})]
& \hspace*{.5cm} &
{\rm if} \hspace*{.2cm} 
\xi_{\rm n}\ge -\xi_{\rm p}, \xi_{\rm n}\ge -1  \\
0
& & {\rm otherwise} \ , 
\end{array}
\right.
\ee
where
\be
\nonumber
\xi_{\tau}=\frac{P^2/4+q^2-(k_{\rm F}^\tau)^2}
{Pq} \ . 
\ee
The mass operators $M_1$ and $M_2$ become 
\be
M_1^{\tau\tau'}(k,\omega) &=&
\frac{1+\delta_{\tau,\tau'}}{2\pi}\sum_{\alpha L}
(2J+1)\int_0^{k^{\tau'}_{\rm F}}
k'^2{\rm d}k'\sin\theta{\rm d}\theta 
G^{\tau\tau'}_{\alpha LL}(q,q,P,\omega+\epsilon^{\tau'}(k')) \ , 
\\[3mm]
M_2^{\tau\tau'}(k,\omega) &=&
\frac{2(1+\delta_{\tau,\tau'})}{\pi^2k}
\sum_{\alpha LL'}(2J+1)
\int\int q{\rm d}q P{\rm d}P \ 
\left[1-n^{\tau'}\left(\sqrt{P^2/2+2q^2-k^2}\right)\right]
\cr
&\times&
\int q'^2{\rm d}q' \langle R^{{\tau}{\tau'}}(q',P)\rangle 
\frac{|G^{\tau\tau'}_{\alpha LL'}(q,q',P,
e_{12}^{\tau\tau'}(q',P))|^2} 
{\omega+\epsilon^{\tau'}\left(\sqrt{P^2/2+2q^2-k^2}\right)
-e_{12}^{\tau\tau'}(q',P)-i\eta} \ . 
\ee
The integrations of $q$ and $P$ in the expression of 
$M_2$ are limited to
\be
q_{{\rm min}} &=& \left\{ 
\begin{array}{llc}
\displaystyle 
\max{\left[0,\ 
k-\frac{1}{2}(k^{\tau}_{\rm F}+k^{\tau'}_{\rm F}),\  
\frac{1}{2}(k^{\tau'}_{\rm F}-k) \right]}
\hspace*{1.4cm} &\ \ {\rm if}& \ k \le k^{\tau}_{\rm F}
\cr\noalign{\vskip 3mm}
\displaystyle
\max{\left[\frac{1}{2}\sqrt{2k^2-2(k^{\tau}_{\rm F})^2+
(k^{\tau}_{\rm F}-k^{\tau'}_{\rm F})^2}, \  
\frac{1}{2}(k+k^{\tau'}_{\rm F}) \right]}
&\ \ {\rm if}& \ k > k^{\tau}_{\rm F}
\end{array}
\right.
\\[3mm]
q_{\rm max} &=&
 k+\frac{1}{2}(k^{\tau}_{\rm F}+k^{\tau'}_{\rm F}) 
\ee
and
\be
P_{{\rm min}} &=& \left\{
\begin{array}{llc}
\displaystyle
\max{\left[
2(k-q),\ 
\sqrt{2k^2+2(k^{\tau'}_{\rm F})^2-4q^2}\right]}
\hspace*{1.2cm} &\ \ {\rm if}& \ q \le 
\frac{1}{2}(k+k^{\tau'}_{\rm F})
\cr\noalign{\vskip 3mm}
\displaystyle
2|q-k|
&\ \ {\rm if}& \ k > \frac{1}{2}(k+k^{\tau'}_{\rm F})
\end{array}
\right.
\\[3mm]
P_{\rm max} &=&
\min{\left[
2(k+q),\ 
(k^{\tau}_{\rm F}
+k^{\tau'}_{\rm F})\right]} \ . 
\ee
The angular averaging of the anti-Pauli operator 
$R^{\tau\tau'}(k_1,k_2) \equiv n^\tau (k_1) n^{\tau'}(k_2)$ can be 
written as 
\par
(i) for $\tau = \tau'$, i.e., neutron-neutron or proton-proton
\be
\langle R^{\tau\tau}(q,P)\rangle &=&
\left\{
\begin{array}{llc}
{\rm min}(1,\eta_{\tau})
& \hspace*{.5cm} &
{\rm if} \hspace*{.2cm} \eta_{\tau}\ge 0  \\
0 & &
{\rm otherwise} \ , 
\end{array}
\right.
\label{e:Apaul1}
\ee
\par
(ii) for $\tau\ne\tau'$, i.e., neutron-proton or proton-neutron
\be
 \langle R^{{\tau}{\tau'}}(q,P)\rangle &=&
\left\{
\begin{array}{llc}
\displaystyle
\frac{1}{2}[{\rm min}(1,\eta_{\rm p})+
{\rm min}(1,\eta_{\rm n})]
& \hspace*{.5cm} &
{\rm if} \hspace*{.2cm}
\eta_{\rm p}\ge -\eta_{\rm n}, \eta_{\rm p}\ge -1  \\
0
& & {\rm otherwise} \ , 
\end{array}
\right.
\label{e:Apaul2}
\ee
where $\eta^{\tau} = -\xi^{\tau}$. 

\section{Results}

We have performed a set of nuclear matter calculations for the 
asymmetric case within the EBHF approximation. Three 
different densities have been selected: $\rho_0 /2$, 
$\rho_0$ and $2\rho_0$, being $\rho_0=0.17 fm^{-3}$  
the saturation density of symmetric nuclear matter. For each 
density the whole range of asymmetry parameter 
( $0 \leq \beta \leq 1$ ) has been spanned. The self-consistent 
solution of the Bethe-Goldstone equation yielding simultaneously 
$G$-matrix and auxiliary potential $U^{\tau}(k)$ needed five 
iterations to reach a satisfactory convergence. The bare potential 
adopted as input in the calculation was the 
Argonne $V_{14}$ \cite{WIRI} with 24 channels up to $L=6$. 
\vskip 0.5 truecm
\subsection{Symmetry Energy}
\vskip 0.5 truecm
In Fig.~4 (left panel) we report the results (symbols) for the 
energy per nucleon $B(\rho,\beta)$, calculated self-consistently 
within the BHF approximation \cite{BKL}. $B(\rho,\beta)$ is plotted 
as a function of $\beta^2$, for three values of density. The 
numerical results lie on a linear fit performed with only the first 
three values of the asymmetry parameter. This proves that the 
empirical parabolic law 
\be
      B(\rho,\beta) = B(\rho,0) + E_{sym}(\rho) \beta^2 \ , 
\label{e:b2law}
\ee
taken from the nuclear mass table can be extended up to the highest 
asymmetry of nuclear matter, in good agreement with our previous BHF 
calculation with the separable Paris potential \cite{LOM}. 
\begin{figure}[ht]
\centerline{\epsfig{figure = 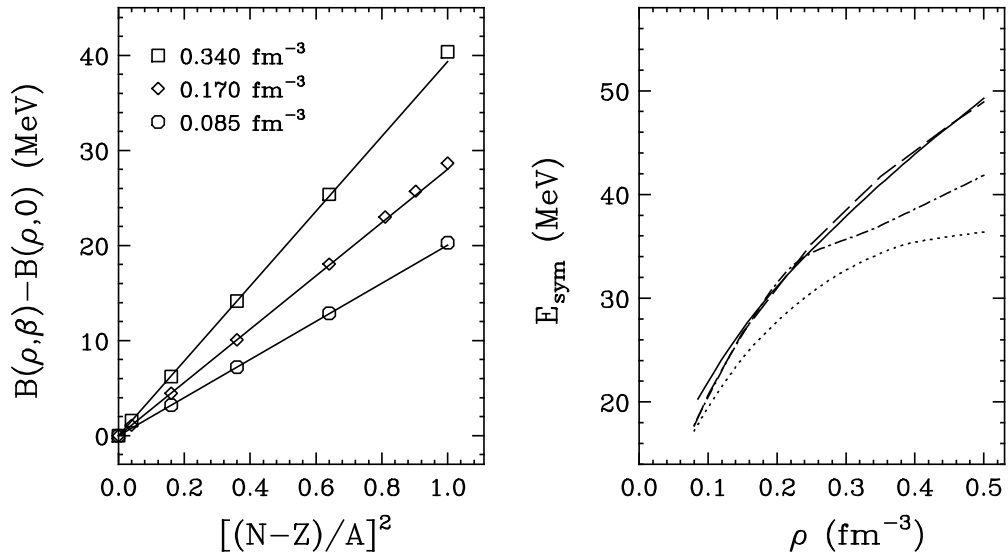,angle=90,width=14cm}}
\caption{\small{
Left panel : Total binding energy per nucleon in the range 
$0\le \beta^2 \le 1$ at three densities as compared with 
the parabolic fits ( straight lines ) obtained from the 
first three values of $\beta$ (0.0, 0.2, 0.4). \ 
Right panel : Density dependence of the symmetry energy of 
the present work ( solid curve ) using Argonne $V_{14}$ as bare 
interaction 
in comparison with other non-relativistic calculations. 
The dashed curve is the result of the lowest order constrained 
variational calculation using Argonne $V_{14}$ as bare 
interaction from 
Ref.[5]. The dotted and dot-dashed curves are the results 
of the variational approach using Argonne $V_{14}$ and 
Argonne $V_{14}$+UVII, respectively, taken from Ref.[3]}} 
\end{figure}\normalsize

Equation (\ref{e:b2law}) can be considered as the $\beta^2$ expansion 
of the binding energy truncated at the lowest order. Only even powers 
of the asymmetry parameter $\beta$ may occur in the expansion for 
charge-independent $NN$ interactions, such as the Argonne $V_{14}$ 
used in the present work. A $\beta^4$ contribution might arise at 
the three hole-line order of the BBG expansion. Unfortunately, no 
such calculation for $B(\rho,\beta)$ has been done yet. However, 
it has been shown recently that the three hole-line contribution to 
the binding energy of symmetric nuclear matter \cite{SBGL} is rather 
small within the continuous choice. Therefore, we do not expect 
a large deviation from the parabolic law after including the three 
hole-line contribution. 
A deviation from the parabolic law could be expected at densities 
higher with respect to those considered in the present 
paper \cite{MODA,HUB,KUO,ENG}.
  
The symmetry energy is defined as 
\be
E_{sym}(\rho) \quad = \quad {1\over 2} \left[ {\partial^2 B(\rho,\beta) 
\over \partial \beta ^2}\right]_{\beta=0}.   
\ee 
Due to the simple $\beta^2$-law the symmetry energy can be 
equivalently calculated as the difference between the binding 
energy of pure neutron matter and symmetric nuclear matter: 
$E_{sym}(\rho) \,=\, B(\rho,1) - B(\rho,0)$, but one would refrain 
from using that recipe at very high density. 
The results of our BHF calculations for $E_{sym}(\rho)$ are depicted by 
the continuous curve in the right panel of Fig.~4. In the same figure, 
we also show the results from the variational approach using 
the same Argonne $V_{14}$ potential \cite{WFF}. 

The systematic disagreement displayed by the two many-body approaches 
has been believed to be a shortcoming of the Brueckner approach in 
view of the fact that the BHF result lies above the variational 
one. However, in Ref.~\cite{WFF} (and similar works), the variational 
expectation value $E_{var}$ of the Hamiltonian is calculated in a 
diagrammatic cluster expansion (FHNC-SOC), which is of course 
truncated to some order. To estimate the convergence of this 
diagrammatic cluster expansion, we plot, in the same figure, the 
results of a lowest-order constrained variational 
calculation~\cite{MODA}, which includes only two-body cluster 
contributions to $E_{var}$. Moreover, the variational trial 
wave function used in Ref.~\cite{WFF} does not contain the 
correlations which arise from 
${\vec{L}}^2$, $\vec{L}^2(\vec{\sigma}_1\cdot\vec{\sigma}_2)$, 
and $(\vec{L}\cdot\vec{S})^2$ terms 
of the nucleon-nucleon potential. Finally, spin-orbit 
correlations are not treated accurately, as discussed in the 
same paper~\cite{WFF}. All these $NN$ correlations are included 
in a self-consistent way in the BHF approach. All the 
above-mentioned approximations could give large uncertainties in 
the calculated expectation value of the energy in the high density 
region. The same discrepancy has also been 
observed in our previous calculations for asymmetric nuclear 
matter \cite{LOM} and also in neutron matter calculations \cite{BAL}. 

From the previous discussion we guess that the nice agreement between 
our calculation and a lowest order constrained variational 
calculation~\cite{MODA} is fortuitous. On the other hand, the 
agreement up to $\rho\sim 0.24$ fm$^{-3}$ with the variational 
calculation including three-body force~\cite{WFF}, also plotted 
in Fig.~4, is hardly understandable.
%%%%%%%%%%%%%%%%%%%%%%%%%%%%%%%%%%%%%%%%

\begin{figure}[ht]
\centerline{\epsfig{figure = 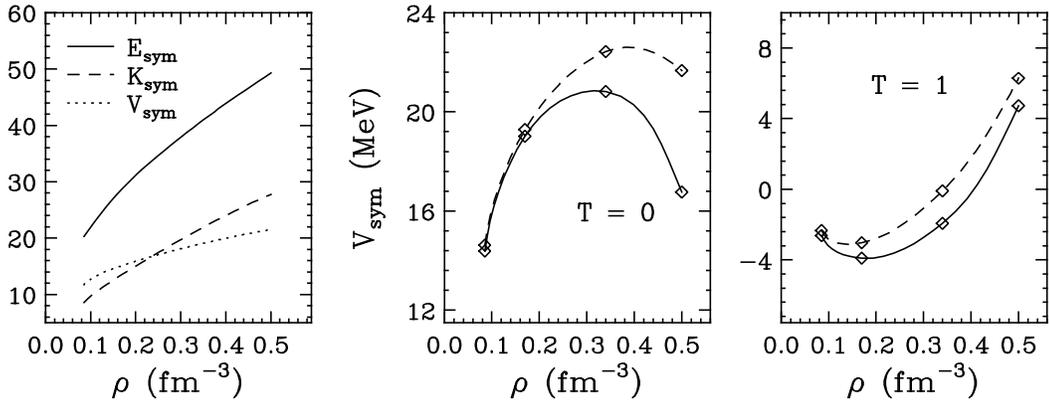,angle=90,width=14cm}}
\caption{\small{Symmetry energy vs density in the BHF approximation. 
Left panel: symmetry energy ($E_{sym}$), kinetic ($K_{sym}$) and 
potential ($V_{sym}$) contributions (only Argonne $V_{14}$). 
Middle panel: potential contribution from isospin $T$=0 channels with 
Argonne $V_{14}$ (solid curve) and separable Paris (dashed curve). 
Right panel: potential contribution from isospin $T$=1 channels with 
Argonne $V_{14}$ (solid curve) and separable Paris (dashed curve). }} 
\end{figure}\normalsize

More recent versions of the $NN$ potential do not provide any 
appreciable difference of the symmetry energy from the present 
calculation except the CD-Bonn potential as discussed in 
Ref.~\cite{ENG1}. All Brueckner calculations predict the symmetry 
energy to increase with the nucleon density and no saturation is 
observed up to $\rho=0.5 fm^{-3}$ at variance with the 
preceeding variational results \cite{WFF}. 
In the relativistic mean-field theory this behavior is easily 
understood in terms of the $\rho$-meson exchange, which leads to 
a repulsive symmetry potential at all densities \cite{BAO,KUO}.  
In order to try to explain what happens in the nonrelativistic case, 
we report in Fig.~5 the different contributions to the symmetry 
energy, plotted as a function of density. The kinetic contribution 
monotonically increases as $\rho^{2\over3}$ according to the free 
Fermi-gas model. In the figure (right two panels), 
the isoscalar and isovector contributions of potential part 
are plotted separately. 
The density dependence of the symmetry energy is dominated at 
high density by the kinetic contribution, where the 
opposite behavior vs density of the potential
contributions from $T=0$ and $T=1$ channels results in a very flat density 
dependence of the symmetry potential.
As already found in the previous paper~\cite{LOM}, the most important 
contribution to the $T=0$ component is due to the deuteron 
$^3 S_1$-$^3 D _1$ coupled channels of the interaction, 
which exhibits a maximum at $ \rho \simeq 0.3 fm^{-3}$. This peak can 
be traced back to the behavior of the two components: the 
attractive $^3 S_1$ channel dominates at low energy 
whereas the repulsive $^3 D_1$ dominates at high energy. Two terms 
compensate each other at the energy $E\simeq 4E_F\simeq 200 MeV$, 
where $E_F$ is the Fermi energy corresponding to 
$ \rho \simeq 0.3 fm^{-3}$.   

\vskip 0.2 truecm
\subsection{Single-particle energy}
\vskip 0.2 truecm
For asymmetric nuclear matter the neutron mass operator $M^n$ is different
from the proton mass operator $M^p$. Moreover, as shown in Figs.~2 
and 3 [see also Eqs.(7) and (8)], both of them can be split into two
components: $M^p=M^{pp}+M^{pn}$ for protons and $M^n=M^{nn}+
M^{np}$ \\
\begin{figure}[ht]
\centerline{\epsfig{figure = 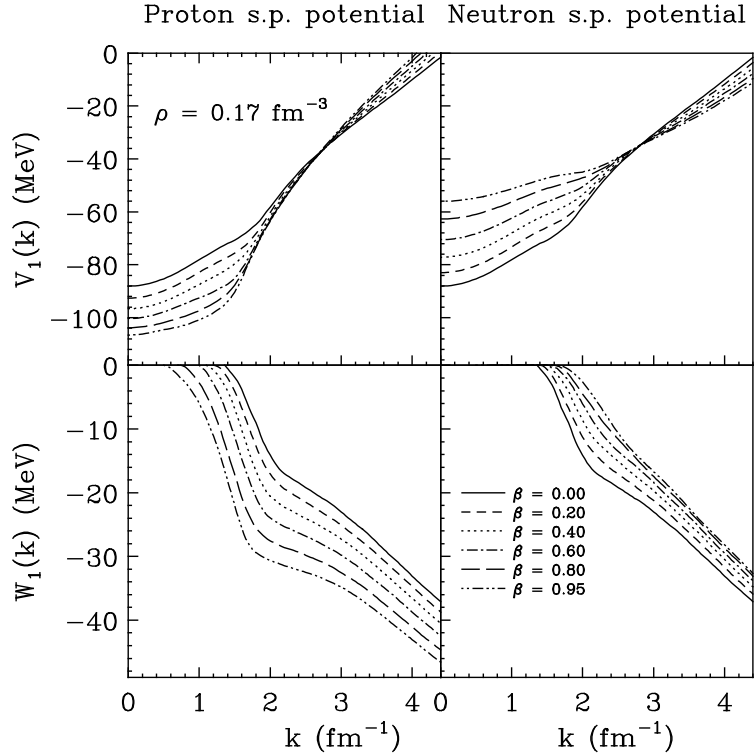,angle=90,width=10cm}}
\caption{\small{Real part (upper panels) and imaginary 
part (lower panels) 
of the first-order single-particle potentials $M_1$ for 
proton (left panel) and neutron (right panel), respectively, as 
a function of momentum for different asymmetry parameters at 
density $\rho = 0.17 fm^{-3}$.}}
\end{figure}\normalsize
\\
for neutrons.  In Fig.~6 the on-shell 
values of the real part (upper panels) of $M_1^{\tau}$ are 
reported as a function of the s.p. momentum, for different values of 
the asymmetry parameter $\beta$ at fixed density $\rho=0.17 fm^{-3}$. 
The proton mean field $V_1^p (k)\equiv \mbox{Re}M_1^p (k)$ becomes 
more attractive, while the neutron mean field 
$V_1^n (k)\equiv \mbox{Re}M_1^n (k)$ becomes more repulsive 
going from symmetric ($\beta=0$) to neutron ($\beta=1$) matter. 
The $\beta$ dependence of $V_1^n$ and $V_1^p$ is almost linear 
and nearly symmetric with respect to their common value at 
$\beta=0$. This result supports from a 
microscopic point of view the validity of the so-called 
Lane potential~\cite{LANE}. 

It is worth noticing that a crossing point occurs for both $V_1^p$ and
$V_1^n$, where the isospin effect on neutron and proton mean fields 
versus $\beta$ is inverted. This behavior of the neutron and proton 
mean field can be understood in terms of phase-space arguments, as 
pointed out Ref.\cite{LOM}. To this end, we write the s.p. potentials 
$V_1^n$ and $V_1^p$ in terms of their components $V_1^{\tau\tau'}$ 
[defined according to Eq.(7)]:
\be
V_1^p(k) \, \simeq \, \frac{1}{2} (1-\beta)\rho \langle G^{pp} \rangle +   
\frac{1}{2} (1+\beta)\rho \langle G^{pn} \rangle 
\\
V_1^n(k) \, \simeq \, \frac{1}{2} (1-\beta)\rho \langle G^{np} \rangle  +   
\frac{1}{2} (1+\beta)\rho \langle G^{nn} \rangle 
\ee
where $\langle G^{pp} \rangle$ is the average value of the real 
part of the matrix $G^{pp}$ in the proton Fermi 
sphere ($|\vec{h}'| \leq k^p_F$), and 
$\langle G^{pn} \rangle$ the average value of the real 
part of the matrix $G^{pn}$ in the neutron Fermi 
sphere ($|\vec{h}'| \leq k^n_F$). $\langle G^{nn} \rangle$ and 
$\langle G^{np} \rangle$ have similar definitions. 
This approximation is suggested by the almost linear dependence 
of $V_1^n$ and $V_1^p$ on $\beta$ and, in fact, 
is numerically fulfilled with a good accuracy (see also Fig.~8). 
The crossing point in momentum space is determined by the occurrence 
of $\langle G^{pp}\rangle = \langle G^{pn}\rangle$ for 
$V_1^p$ and $\langle G^{np}\rangle = \langle G^{nn}\rangle$ for 
$V_1^n$ at a certain value of the momentum which does not 
depend upon $\beta$.
A signature of the inversion of the isospin effect at the crossing 
point could be found in those collective observables measured in heavy 
ion collisions which are sensitive to the momentum dependence of 
the mean field.   

\begin{figure}[ht]
\centerline{\epsfig{figure = 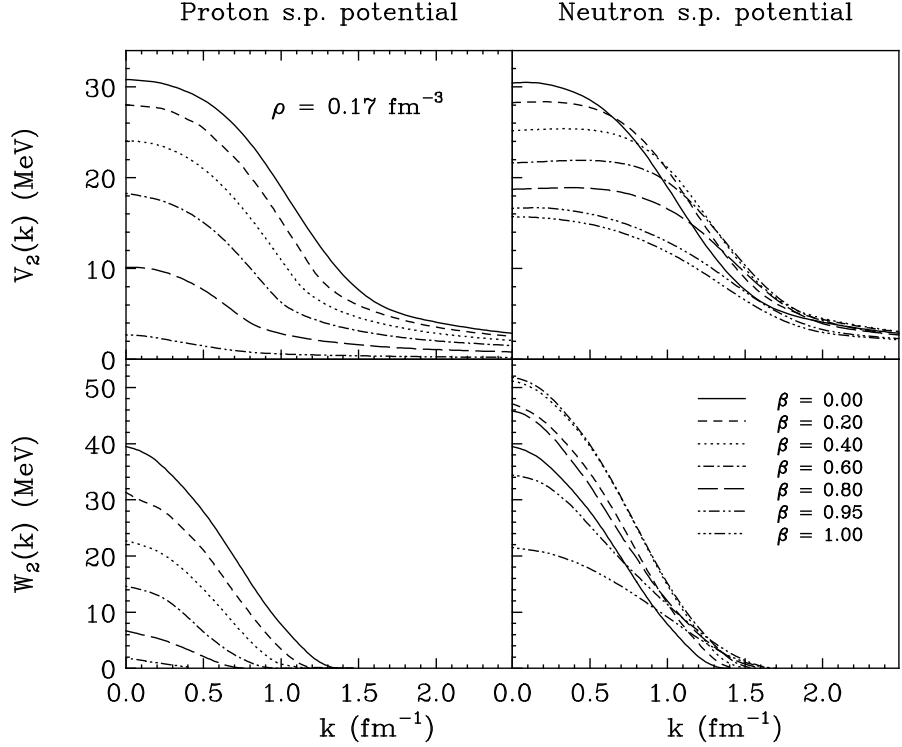,angle=90,width=12cm}}
\caption{\small{Real part (upper panels) and imaginary part (lower panels) 
of the second-order single-particle potentials $M_2$ for 
proton (left panel) and neutron (right panel), respectively, as 
a function of momentum for different asymmetry parameters at 
density $\rho = 0.17 fm^{-3}$. }}
\end{figure}\normalsize
\begin{figure}[ht]
\centerline{\epsfig{figure = 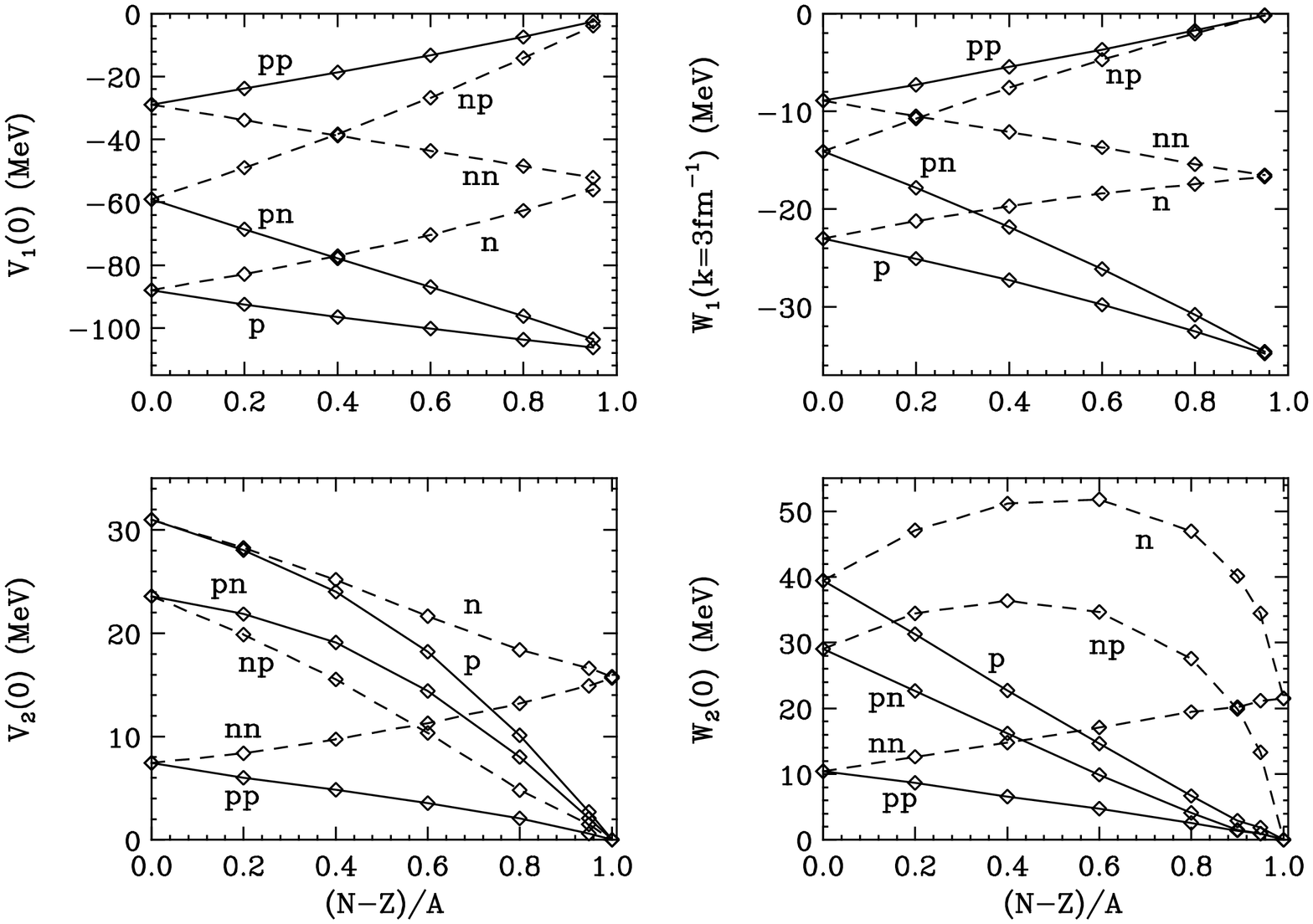,angle=90,width=14cm}}
\caption{\small{Different components of the first-order (upper panels) 
and the second-order (lower panels) single-particle potentials 
at density $\rho=0.17 fm^{-3}$ for suitable values of momentum 
versus asymmetry parameter. 
For the real part of $M_1$ and $M_2$, and the Imaginary part of $M_2$, 
the momentum is $k=0$, while for the real part of $M_1$ the momentum 
is $k=3 fm^{-1}$. }} 
\end{figure}\normalsize

The imaginary part $W_1^{\tau}$ of the mass operator $M_1^{\tau}$ 
is due to the virtual collisions of a single nucleon with a neutron or 
a proton of the background, promoting it to a particle state. 
$W_1^{\tau}$ is vanishing below the Fermi momentum $k_F^{\tau}$ due 
to the Pauli blocking. 
It is worth noticing that reducing the proton Fermi momentum 
implies a less Pauli blocking for protons. This means that high 
asymmetric nuclear matter is less transparent to the proton 
propagation. 

The second-order terms of the on-shell mass operator $M_2^{\tau}$ are 
plotted in Fig.~7. The real part $V_2^{\tau}$ (upper panels) gives the 
contribution to the mean field due to the coupling of the 
single-particle motion with the ground state particle-hole excitations. 
As is well known, $V_2^{\tau}$ is repulsive and reduces to a large 
extent the pure BHF mean field $V_1^{\tau}$ which is too attractive 
compared with the phenomenological optical potential \cite{GRAN}.

The imaginary part $W_2^{\tau}$ plays a role complementary to 
$W_1^{\tau}$: it describes the virtual collisions of a single nucleon 
of the background with an excited neutron or proton, making it to decay 
into a hole state. $W_2^{\tau}$ is vanishing above the Fermi momentum 
$k_F^{\tau}$. 

In order to focus on only the isospin dependence, we plot in Fig.~8 
the mass operator as a function of $\beta$ at $k=0fm^{-1}$ except for 
the imaginary part of $M_1$ for which a value of $k$ above the Fermi 
momentum has to be taken. 

The first-order contribution has the linear behavior for the real 
part as well as for the imaginary part as expected from 
phase-space arguments. The slope of $|W_1^{pn}|$ is 
more pronounced than that of $|W_1^{np}|$ since the neutron 
particle-hole excitations coupled to a proton in a particle state 
are more favored than the proton particle-hole excitations 
coupled to a neutron (see also Fig.~6).
 
The isospin dependence of the second-order contribution 
$M_2^{\tau}$ is affected by the coupling between the nucleon 
hole stats and particle-hole excitations [see the 
bubble in Fig.3 and Eq.(\ref{eq:M2})], 
which yields a nonlinear variation of the mixed components 
$M_2^{pn}$ and $M_2^{np}$ vs $\beta$. 
The nonlinearity is much more sizeable for $V_2^{pn}$ and 
$W_2^{np}$, which can be easily explained as a phase-space effect 
as well, {\it i.e.,} of the interplay between the neutron and 
proton phase-spaces as increasing neutron excess. 

\vskip 0.2 truecm
\subsection{Fermi Energy and Hugenholtz-Van Hove theorem}
\vskip 0.2 truecm
The EBHF approximation basically relies on the Landau definition of 
quasi-particle energy as showed in sec.I, whose relation to the 
Brueckner theory has been well established \cite{BROW}. Study of the 
HVH theorem within the EBHF approximation could provide an additional 
support to a proper definition of the quasi-particle energy and, 
at the same time, a more realistic evaluation of the Fermi energy.

Strictly speaking, the HVH theorem concerns only symmetric nuclear 
matter at saturation point ($P\,=\,0$), and it states that the energy 
per nucleon must be exactly equal to the Fermi energy. 
In the case of asymmetric nuclear matter (two-component system) at zero 
temperature, the HVH theorem can be generalized via the thermodynamic 
relation
\be
\frac{E(\rho,\beta)}{A}+\frac{P(\rho,\beta)}{\rho}
= Y^p E_F^{\rm p}(\rho,\beta)
+ Y^n E_F^{\rm n}(\rho,\beta) \ , 
\label{e:HVH}
\ee
$P(\rho,\beta)$ being the pressure, $Y^p=\rho_p/\rho$ and 
$Y^n=\rho_n/\rho$ the proton and neutron fractions, respectively. 
The Fermi energy is calculated from the quasi-particle 
energy spectrum at Fermi surface according to Eq.~(6). \\[.2cm]
%%%%%%%%%%%%%%%%%%%%%%%%%%%%%%%%%%%%%%%%%%%%%%%%%%%%%%%%%%%%%%%%%%%%%%%%
%%                  TABLE I
%%%%%%%%%%%%%%%%%%%%%%%%%%%%%%%%%%%%%%%%%%%%%%%%%%%%%%%%%%%%%%%%%%%%%%%%
\footnotesize{ TAB.~I \ \ As a function of the asymmetry 
parameter (first comumn) are reported (in MeV) the physical 
quantities involved in the Hugenholtz-Van Hove theorem: 
pressure over $\rho$ (second column), energy per 
nucleon (third column) and ``weighted'' chemical potentials of 
asymmetric nuclear matter 
in different approximations, as discussed in the text. 
$Y^p = Z/A$ and $Y^n = N/A$ are the proton and neutron fractions, 
respectively. The total density is $\rho=0.17fm^{-3}$. } 
\normalsize
\begin{center}\small
\begin{tabular}{|c|c||c|c|c||c|c|c||c|c|c||c|}
\hline
$\beta$ & $P/\rho$ & $E/A$ & $P/\rho+E/A$ & 
$Y^{\rm p} E_F^p + Y^{\rm n} E_F^n$ & 
$Y^{\rm p} E_F^p + Y^{\rm n} E_F^n$ & 
$Y^{\rm p} E_F^p + Y^{\rm n} E_F^n$ 
\\
 &  &  &  & BHF  & BHF+$M_2$ & EBHF 
\\ \hline\hline
0.0 & $-5.02$ & $-15.92$ & $-20.94$ & $-34.27$ 
& $-28.50$ & $-19.28$ 
\\ \hline
0.2   & $-4.40$ & $-14.73$ & $-19.13$ & $-32.29$
& $-26.43$ & $-17.35$
\\ \hline
0.4   & $-3.27$ & $-11.36$ & $-14.63$ & $-26.28$ 
& $-20.74$ & $-12.42$ 
\\ \hline
0.6   & $0.08$  & $-5.75$ & $-5.67$ & $-16.56$ 
& $-11.44$ & $-4.41$ 
\\ \hline
0.8   & $3.76$  & $2.24$  & 6.00    & $-2.40$ 
& $2.07$ & 7.42 
\\ \hline
1.0 & $8.91$ & $12.83$ & $21.74$ & 16.28 
& 19.67 & 22.39
\\ \hline
\end{tabular}
\end{center} 
\normalsize\vspace*{.5cm}
In Tab.~I it is numerically shown to what extent the HVH 
theorem is fulfilled by the EBHF approximation. 
The pressure has been calculated using the relation 
$P(\rho,\beta) = \rho^2[\partial E_A(\rho,\beta)/\partial\rho]$, 
where $E_A (\rho,\beta) \equiv E(\rho,\beta)/A$ being the 
energy per nucleon.  
In the forth column the left-hand side of Eq.~(\ref{e:HVH}) is 
calculated for several asymmetries (density fixed at 
$\rho \,=\,0.17fm^{-3})$. One would notice that, despite the fact 
that the total density is fixed at the empirical saturation value, 
our calculated saturation point lies at higher density, because, 
as is well known, Brueckner theory with two-body force misses the 
empirical saturation point. The last three columns 
provide different approximations for the right-hand side of 
Eq.~(\ref{e:HVH}). The pure BHF approximation by itself is far from 
fulfilling the HVH theorem. Including the unrenormalized ground-state 
correlations (indicated by BHF$+M_2$ in the table), where the Fermi 
energy is calculated according to Eq.~(9), provides some improvement 
but it is not enough to fulfill the HVH theorem. One needs to include 
both the rearrangement and the renormalized 
contributions (EBHF) if a satisfactory agreement within less 
than $10\%$ 
is to be attained (last column of Tab.~I). 
This result is in keeping with the uncertainty in the 
calculation of the pressure because the binding energy curve is 
rather flat as a function of density.

\section{Applications}

\vskip 0.5 truecm
\subsection{Effective mass}
\vskip 0.5 truecm
The effective mass incorporates the non-local part of the mean field 
which makes the local part less attractive for a nucleon travelling 
with momentum $ k > 0$. It is defined as
\be
{ m^*_{\tau}(k) \over m} \, = \, {k\over m} 
\left( {dE^{\tau}(k)\over{dk}} \right)^{-1}.
\ee
The momentum dependence of $m^*$ is characterized by the wide bump 
inside the Fermi sphere due to the high probability amplitude for 
particle-hole excitations near the Fermi surface \cite{MAHA0}. The 
effect of correlations is a flattening of the slope of the mean field 
around the Fermi energy, which implies an enhancement of the effective 
mass at $k_F$ with respect to BHF value \cite{ZUO1,ZUO2}. This 
result is shown in Fig.~9, where an increase from 0.8 to 
0.92 is observed for symmetric nuclear matter at the empirical 
saturation density. Also shown in the figure is the isospin dependence 
of the neutron (upper curve) and proton (lower curve) effective 
masses. In both the BHF and EBHF calculations, 
$m^*_n$ increases and $m^*_p$ decreases as increasing $\beta$. 
Compared to the BHF approximation, the corrections of EBHF 
shift $m^*_n$ and $m^*_p$ to higher values, a feature 
which can be traced to the depletions of the proton and 
neutron Fermi surfaces due to the ground-state correlations. 
The value of $m^*_p$ calculated from EBHF approaches its BHF 
value as increasing $\beta$ since the correlations become smaller. 
\begin{figure}[ht]
\centerline{\epsfig{figure = 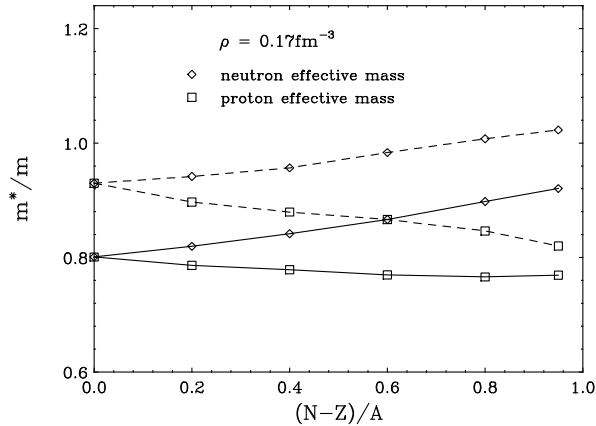,angle=90,width=8cm}}
\caption{\small{Proton and neutron effective masses versus asymmetry 
parameter at density $\rho=0.17 fm^{-3}$. The solid curves are results 
from the pure BHF calculation, while the dashed curves are calculated 
from the EBHF (including the renormalization contributions). }}
\end{figure}\normalsize

\vskip 0.5 truecm 
\subsection{Mean free path}

Information on the in-medium cross section or, equivalently, on 
the mean free path of a nucleon travelling inside a nuclear medium 
can be obtained from the transparency of a nucleus measured 
in $(e,e' p)$ reactions \cite{PAND} and, in general, from 
nucleon-induced reactions at low energy \cite{AVRI}. The underlying 
assumption is that the behavior of a nucleon located at the 
position $\vec{r}$ in a nucleus is the same as a nucleon in nuclear 
matter at density $\rho(\vec{r})$. Such an assumption is the well 
known local-density approximation (LDA) \cite{SCHU}. 
The mean free path is intimately related to the imaginary part of 
the optical potential or, equivalently, to the imaginary part of 
the mean field. The latter comes from the collisions 
of a single nucleon with the background of neutrons and 
protons: a nucleon with momentum $ k \ge k_F $ can collide with a 
neutron or proton of its Fermi sea and promote it to a particle 
state, or a nucleon with momentum  $k \le k_F$ interacting with 
an excited neutron or proton can make it decay into a hole 
state. The first process is related to the imaginary part of $M_1$, 
the second one to the imaginary part of $M_2$, 
both of which have been plotted in the lower panels of Figs.~6 and 7. 
But, in the case of asymmetric matter the collisions between like 
and unlike nucleons yield contributions to the mean field which 
are very different. 

\begin{figure}[ht]
\centerline{\epsfig{figure = 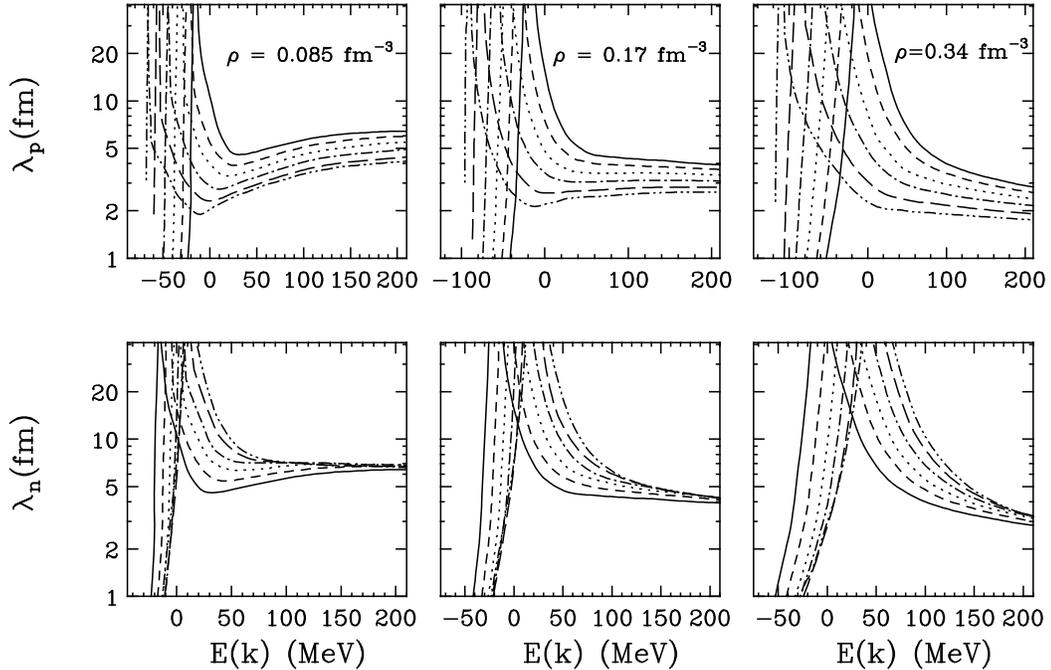,angle=90,width=14cm}}
\caption{\small{The energy dependence of proton (upper panels) and 
neutron (lower panels) mean free paths for different 
asymmetry parameters at three densities $\rho = 0.085$fm$^{-3}$, 
$\rho = 0.17$fm$^{-3}$, and $\rho = 0.34$fm$^{-3}$. Only EBHF results 
are reported. The different kinds of lines correspond to different 
asymmetry parameters with the same notation as in Fig.6.}}
\end{figure}

The mean free path $\lambda_{\tau}$ is given by 
\be
\nonumber
\lambda_{\tau}(E) 
\, = \, {\hbar^2 k(E) \over{2 {\tilde m_{\tau}}}}
{1 \over {|{\rm Im}M_{\tau}(k(E),E)|}} \ , 
\ee 
where $\tilde m_{\tau}$ is the so-called $k$-mass, and 
$E$ is the single-particle energy \cite{MAHA2}. 
In Fig.~10 the proton (upper panels) and neutron (lower panels) 
mean free paths calculated within the EBHF approximation are 
shown for three values of the total density. 
In each panel the values of $\lambda_{\tau}$ for several asymmetries 
are plotted as a function of single-particle energy. The most relevant 
effect of the isospin asymmetry is the increasing deviation from 
the symmetric values (solid lines), upward 
for $\lambda_n$ and downward $\lambda_p$, as increasing asymmetry. 
The nonvanishing values of neutron and proton mean free paths 
below their respective Fermi energies are effects of ground-state 
correlations which prevents a full occupancy of the Fermi spheres. 
Comparing with the BHF calculation it turns out that the correlation 
effects tend to rise the asymptotic value of the mean free path from 
about $3$fm up to about $4$fm at the saturation density of symmetric 
nuclear matter \cite{ZUO2}. 
\vspace*{1cm}
\begin{figure}[ht]
\centerline{\epsfig{figure = 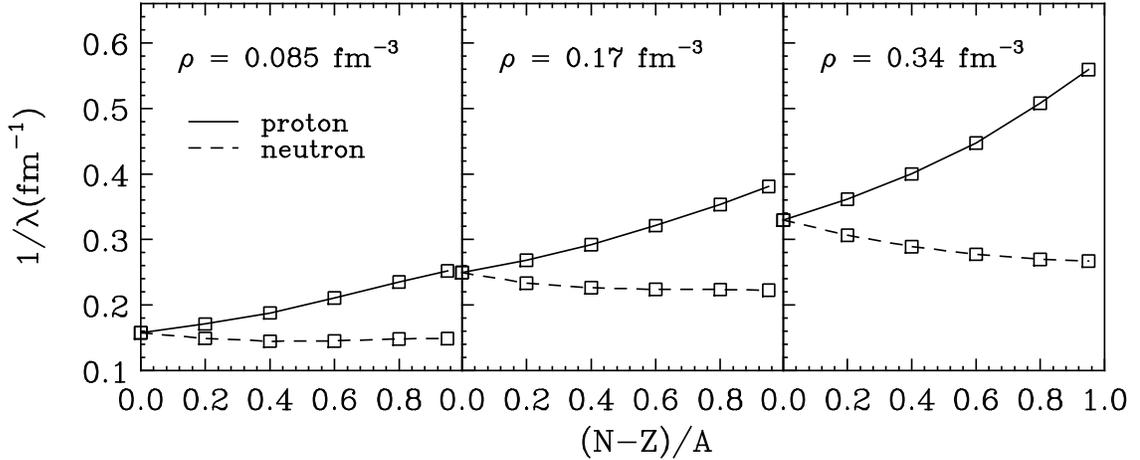,angle=90,width=15cm}}
\caption{\small{Proton and neutron inverse mean free paths versus 
asymmetry parameter for three densities 
$\rho = 0.085$fm$^{-3}$, $\rho = 0.17$fm$^{-3}$, and 
$\rho = 0.34$fm$^{-3}$ at a fixed single-particle energy 
$E^{\tau}(k)=180$MeV from the EBHF calculation. }}
\end{figure}
 
In Fig.~11 it is shown how the isospin dependence of the inverse 
$\lambda$ develops as increasing nuclear matter densities at a 
fixed value of the energy. Except for very small asymmetries 
the shift of $\lambda_p$ and $\lambda_n$ is not symmetric with 
respect to their common value at $\beta = 0$. At any density the slope 
of the neutron inverse $\lambda$ is less than the proton one. This 
effect can be traced to the reduction, as increasing neutron excess, 
of the proton particle-hole excitations contributing to the neutron 
optical potential (see Figs.~6 and 7). Moreover, the EBHF $\lambda_n$ 
seems to reach the asymptotic value of pure neutron matter much faster 
than in the uncorrelated case.

The most striking effect of the isospin-asymmetry is the sizeable 
reduction of the proton mean free path at high asymmetry. Accordingly, 
the nuclear surface would become more transparent to neutrons than 
protons in nucleon-induced reactions on nuclei near the neutron 
drip-line. This effect would be more pronounced at higher density 
as shown in Fig.~11.    

\vskip 0.5 truecm
\subsection{Proton fraction in $\beta$-equilibrium matter}
\vskip 0.5 truecm

The core of a neutron star is expected to be formed by an uncharged 
mixture of neutrons, protons, electrons and muons in equilibrium with 
respect to the weak interactions ($\beta$-stable matter). 
The concentrations of different particles are then obtained under 
the requirements 
\be
  \mu_n - \mu_p = \mu_e, ~~~~~~~~~~~~~~~~~~
  \mu_\mu = \mu_e.
\label{eq:chem}
\ee
\be
            \rho_p  =  \rho_{e}  +  \rho_{\mu} \ . 
\label{eq:charge}
\ee
The difference between the neutron and proton chemical potentials 
can be expressed as 
\be
 \mu_n - \mu_p = - {{\partial B}\over{\partial Y^p}}\Bigg|_{\rho} =
           2 {{\partial B}\over{\partial \beta}}\Bigg|_{\rho} \ . 
\label{eq:muhat}
\ee
In the parabolic approximation, Eq.~(\ref{e:b2law}), for the energy per 
particle of asymmetric nuclear matter, one has
\be
 \mu_n - \mu_p  = 4 E_{sym}(\rho) (1-2Y^p) \ . 
\label{eq:muhat1}
\ee

Therefore, the composition of $\beta$-stable matter, and in 
particular, the proton fraction $Y^p$  present at a given density, 
is strongly dependent on the nuclear symmetry energy. 
The proton fraction plays also a crucial role in the thermal 
evolution of neutron stars. In fact, if the proton fraction in the 
core of a neutron star, is above a critical value $Y^p_{Urca}$, 
the so-called direct Urca processes can occur \cite{LATT}. 
If they occur, the direct Urca processes enhance the neutrino 
emission and neutron star cooling rate by a large factor compared 
to the standard cooling scenario. 
The critical proton fraction has been estimated \cite{LATT} to be 
in the range 11 -- 15\%. 
In a recent paper \cite{BAL}, based on microscopic EOS of 
dense matter, it has been found that the onset of the direct Urca 
processes occurs at densities $\rho > 0.54$--0.65 $fm^{-3}$, 
depending on the nuclear interaction used to get the 
EOS ( see Ref.~\cite{BAL} for more details ).

In Tab.~II, we report our present calculations of the proton fraction 
$Y^p(\rho)$ for $\beta$-stable matter in comparison with the one 
obtained with the separable Paris ( see Ref.~\cite{LOM} ) and 
variational calculation of Ref.~\cite{WIRI} using the Argonne $V_{14}$ 
potential plus the Urbana model ( UVII ) three-body force. In the 
calculations reported in Tab.~II, muons have been not included. 
\\[.2cm]
%%%%%%%%%%%%%%%%%%%%%%%%%%%%%%%%%%%%%%%%%%%%%%%%%%%%%%%%%%%%%%%%%%%%%%%%
%%                  TABLE II
%%%%%%%%%%%%%%%%%%%%%%%%%%%%%%%%%%%%%%%%%%%%%%%%%%%%%%%%%%%%%%%%%%%%%%%%
\footnotesize{Tab.~II \ \ Proton fraction in $\beta$-stable nuclear 
matter (no muons) versus the total baryonic density from different 
forces. The values reported are $10^2 Y$. The results in the second 
column are taken from Ref.~\cite{LOM}. Those in the third column 
have been given by A. Fabrocini (private communication).}
\normalsize
\begin{center}
\begin{tabular}{|c|c|c|c|c|} \hline\hline
~~~~~~~$\rho_B$~~~~~~~  & ~~~~~Paris~~~~~ & ~~AV14+UVII~~ 
& ~~~present~~~ \\ \hline
 0.038 &  2.75  &  ---       &         \\
 0.076 &  2.80  &  1.85      &  2.40   \\
 0.11  &  3.09  &  2.48      &  2.74   \\
 0.14  &  3.48  &  2.96      &  3.03   \\
 0.17  &  3.70  &  3.37      &  3.32   \\
 0.20  &  4.10  &  3.74      &  3.50   \\
 0.30  &  4.90  &  3.67      &  4.07   \\
 0.40  &  5.79  &  3.56      &  4.57   \\
 0.50  &   ---  &  3.63      &  5.01   \\ \hline \hline
\end{tabular}
\end{center} 
%%%%%%%%%%%%%%%%%%%%%%%%%%%%%%%%%%%%%%%%%%%%%%%%%%%%%%%%%%%%%%%%%%
Our purpose, in the present paper, is not an accurate determination
of the proton fraction in dense stellar matter.
Here, we aim to study how the inclusion of contributions beyond 
the BHF to the chemical potentials could alter the proton fraction 
in $\beta$-stable matter. In fact, to solve the $\beta$-equilibrium 
conditions (\ref{eq:chem}) and (\ref{eq:charge}), the shift between 
neutron and proton chemical potentials $\hat\mu \equiv \mu_n - \mu_p$ 
has to be evaluated. In Tab.~III the neutron and proton chemical 
potentials and their difference $\hat\mu$, are reported for the 
different approximations used in the present work. From the results 
reported in Tab.~III we see that the chemical potential, approximated 
by the Fermi energy, in the EBHF are noticeably affected by the 
rearrangement and renormalization contributions.  
However, their difference and consequently the proton fraction is 
almost unchanged with respect to the 
BHF approximation. The EBHF approximation provides  neutron and proton 
Fermi energies, which are in better agreement with the empirical values 
extracted from the mass table of atomic nuclei \cite{MAHA1} than the 
BHF approximation \cite{LOM}.
\\[5cm]
%%%%%%%%%%%%%%%%%%%%%%%%%%%%%%%%%%%%%%%%%%%%%%%%%%%%%%%%%%%%
%%                  TABLE III
%%%%%%%%%%%%%%%%%%%%%%%%%%%%%%%%%%%%%%%%%%%%%%%%%%%%%%%%%%%
\footnotesize{Tab.~III \ \ Proton and neutron chemical 
potentials (Fermi energies) calculated in different 
approximations and compared with the symmetry energy. 
The results corresponding to four values of asymmetry 
parameter $\beta$ for each of the three densities are reported. }
\\
\begin{center}
\footnotesize
\begin{tabular}{|c|c||c|c|c||c|c|c||c|c|c||c|}
\hline
$\rho$  & $\beta$ & 
$\mu^p$ & $\mu^n$ & $\hat\mu$ & 
$\mu^p$ & $\mu^n$ & $\hat\mu$ & 
$\mu^p$ & $\mu^n$ & $\hat\mu$ & 
$4\beta E_{sym}$ \\ \hline
$fm^{-3}$ & & \multicolumn{3}{|c||}{BHF} & \multicolumn{3}{|c||}
{BHF+$M_2$} 
& \multicolumn{3}{|c||}{EBHF} & \\ \hline\hline
& 0.2 & $-33.99$ & $-18.05$ & 15.94 & $-30.44$ & $-14.18$ &
16.26 & $-23.00$ & $-6.92$ & 16.08 & 16.22
\\ \cline{2-12}
0.085 & 0.4 & $-42.98$ & $-10.77$ & 32.20 & $-39.75$ & $-7.01$ &
32.74 & $-32.60$ & $-0.16$ & 32.44 & 32.45
\\ \cline{2-12}
 & 0.6 & $-51.71$ & $-3.31$ & 48.40 & $-49.26$ & $0.15$ &
49.41 & $-43.24$ & $5.67$ & 48.91 & 48.67
\\ \cline{2-12}
 & 0.8 & $-61.32$ & $3.68$ & 65.00 & $-59.63$ & $6.58$ &
66.21 & $-54.85$ & $10.49$ & 65.34 & 64.90
\\ \hline \hline
& 0.2 & $-45.94$ & $-23.19$ & $22.75$ & $-40.45$ & $-17.08$ & 
23.37 & $-31.01$ & $-8.24$ & 22.77 & 23.00
\\ \cline{2-12}
0.170 & 0.4 & $-58.08$ & $-12.65$ & $45.43$ & $-53.46$ & $-6.72$ &
46.74 & $-44.30$ & $1.25$ & 45.55 & 46.00
\\ \cline{2-12}
& 0.6 & $-71.73$ & $-2.77$ & $68.96$ & $-68.10$ & $2.72$ &
70.82 & $-59.60$ & $9.39$ & 68.99 & 69.00
\\ \cline{2-12}
& 0.8 & $-86.22$ & $6.91$ & $93.13$ & $-83.99$ & $11.63$ &
95.62 & $-76.42$ & $16.73$ & 93.15 & 92.00
\\ \hline \hline
& 0.2 & $-47.53$ & $-16.11$ & 31.42 & $-38.89$ & $-5.75$ &
33.14 & $-24.04$ & $7.12$ & 31.16 & 32.30
\\ \cline{2-12}
0.340 & 0.4 & $-64.84$ & $-1.79$ & 63.05 & $-57.48$ & $8.88$ &
66.36 & $-42.31$ & $20.22$ & 62.53 & 64.59
\\ \cline{2-12}
 & 0.6 & $-82.75$ & $12.21$ & 94.96 & $-77.17$ & $22.85$ &
900.02 & $-62.44$ & $32.35$ & 94.79 & 96.89
\\ \cline{2-12}
 & 0.8 & $-103.15$ & $25.48$ & 128.63 & $-99.87$ & $35.52$ &
135.39 & $-86.32$ & $42.95$ & 129.27 & 129.18
\\ \hline
\end{tabular}
\end{center}
\normalsize
\vskip 0.5 truecm
\section{Conclusions}
\vskip 0.5 truecm
In this paper we have reported the study of asymmetric nuclear matter 
within the Brueckner-Bethe-Goldstone approach. The isospin effect on 
the equation of state has been investigated by performing a set of 
calculations at the two hole-line level of the BBG expansion for 
the energy per particle $B(\rho,\beta)$. The Bethe-Goldstone equation 
has been solved with the Argonne $V_{14}$ interaction. The continuous 
choice has been adopted for the auxiliary potential since it makes 
the convergence of the hole-line expansion faster than the gap 
choice \cite{SBGL}. 
Ranging the asymmetry parameter from $\beta=0$ (symmetric 
nuclear matter) to $\beta=1$ (pure neutron matter) it was possible 
to check that $B(\rho,\beta)$ exhibits a linear dependence on 
$\beta^2$ for baryonic densities as large as at least two times 
the saturation density. This result confirms the 
empirical law introduced in the mass formula of atomic nuclei 
and also extends its validity up to the highest asymmetries. 
As a consequence, the entire isospin effect is incorporated in 
the symmetry energy. The calculation of the symmetry energy in 
the BHF approximation shows a monotonic increase as a function of 
baryonic density. 
Its value calculated at the saturation density is about 
28.7MeV, in agreement with the empirical one. The comparison with 
the variational prediction is made rather difficult due to the 
contradictory results still existing in this approach. 
An accurate determination of the symmetry energy is required 
for dynamical simulations of collisions between neutron-rich nuclei, 
where the collective 
observables including collective flows, balance energy and other 
quantities are expected to be sensitive to the isospin degree 
of freedom \cite{BAO,DITO}. 
Ground state correlations were included in the mass operator 
up to the four hole-line order contributions. 
Their effect on the single particle properties has been investigated. 
The first-order contribution to the mass operator displays a 
linear dependence on the asymmetry parameter confirming a 
long-standing analysis by Lane \cite{LANE}.  
A new effect of the isospin degree of freedom appears when the 
ground-state correlations induced by the second-order contribution 
are introduced in the mass operator. That is a nonlinear effect 
due to the particle-hole excitations of, say, protons induced 
by the propagation of a neutron in the nuclear medium. This new 
feature affects the isospin dependence of single-particle properties 
such as mean field, effective mass and mean free path.  
Along with the symmetry energy the heavy ion collisions with 
asymmetric nuclei could also probe the isospin dependence of 
mean free path and effective mass, which play also an important 
role in the collision dynamics. 

The EBHF approximation for asymmetric matter results  
in a satisfactory fulfillment of the Hugenholtz-Van Hove theorem in 
all asymmetry range $0\leq \beta \leq 1$. 
This property makes us more confident 
of the hole-line expansion of the mass operator for calculating the 
single-particle properties including the Fermi energy. 
We found that the neutron and proton chemical potentials  
are largely affected by contributions beyond the BHF approximation. 
This could have far reaching consequences for the physics of the 
neutron star crust. In fact, the proton chemical potential in 
asymmetric nuclear matter is a very important ingredient in locating 
the inner boundary of the neutron star crust. 
However the difference $\hat\mu \equiv \mu_n - \mu_p$, and 
consequently the proton fraction in $\beta$-stable matter, is almost 
unchanged in the EBHF approximation with respect to the BHF 
approximation.

%%%%%%%%%%%%%%%%%%%%%%%%%%%%%%%    
\section*{Acknowledgment}
%%%%%%%%%%%%%%%%%%%%%%%%%%%%%%%
The authors are indebted to Professor A. Fabrocini for 
valuable discussions.

%%-----------------------------------------------------------------------

\end{document}